\documentclass[aps,prc,nofootinbib,showpacs]{revtex4}

\usepackage{epsfig}
\usepackage{graphicx}

\begin{document}
\title{Polarization observables in the processes $p+p\to \Theta^+ +\Sigma^+$ and $n+p\to \Theta^+ +\Lambda^0$, for any spin and parity of the $\Theta^+$-hyperon in the threshold region.}
\author{Michail P. Rekalo }
\affiliation{National Science Centre - Kharkov Institute of 
Physics and Technology,\\ Akademicheskaya 1, 61108 Kharkov,
Ukraine}
\author{Egle Tomasi-Gustafsson}
\affiliation{\it DAPNIA/SPhN, CEA/Saclay, 91191 Gif-sur-Yvette Cedex, 
France}
\date{\today}


\begin{abstract}
Using the symmetry properties of the strong interaction, such as the Pauli principle, the P-invariance, the conservation of the total angular momentum and isotopic invariance, we establish the spin structure of the threshold matrix elements for the processes $p+p\to \Theta^+ +\Sigma^+$ and $n+p\to \Theta^+ +\Lambda^0$, in a model independent way, which applies to any spin and parity of the $\Theta^+$-hyperon 
in the near threshold region. We predict the double spin observables for these processes, such as the dependence of the differential cross section on the polarizations of the colliding nucleons, and the coefficients of polarization transfer from a nucleon beam or target to the produced $\Sigma^+$ or $\Lambda^0$ hyperon. We prove that these observables are sensitive to the P-parity of the  $\Theta^+$ baryon, for any value of its spin. As an example of dynamical considerations, we analyzed these reactions in the framework of K-meson exchange.
\end{abstract}
\maketitle
\section{Introduction}

The quantum numbers of the possible pentaquark state $\Theta(1540)$ is object of intensive experimental and theoretical considerations. In particular, the P-parity of this baryon is important in order to disentangle different models \cite{Di97,Ja03,St03,Ho03,Sa03,Zh03a}. 

In principle, specific polarization phenomena in different reactions, such as $\gamma+N\to \Theta^++\overline{K}$ \cite{Re04a}, $\pi+N\to \Theta^++\overline{K}$ and   
 $K+N\to \pi^++\Theta^+$ \cite{Bo59,Bi58},  $p+p\to \Theta^++\Sigma^+$ \cite{Th03,Ha03}, $n+p\to \Theta^++\Lambda^0$ \cite{Re04b,Uz04b}, $p+p\to\pi^+ + \Lambda^0+\Theta^+$ \cite{Re04a,Pa99} ...,  can constitute the tool for an adequate and model independent way for the determination of the P-parity of the $\Theta$-hyperon. However, the above listed reactions can be used only if one knows some other quantum numbers of $\Theta$, such as the spin, or, in some cases, the isospin. In general, the mass and the total width do not eneter in such considerations. In the reactions involving the $K$-meson (in initial or in final states) the P-parity of the $K$ meson has to be known. A typical assumption is that the K-meson is pseudoscalar, following the quark model, but, up to now, there are only indirect experimental indications. We may remind that, in the beginning of photo and electroproduction studies, both values were considered in the interpretation of the data, \cite{Fa61}. This problem is crucial in the context of the $\Theta$-classification, as the main $\Theta^+$-decay, $\Theta^+\to NK$, being a strong decay, is controlled by the kaon P-parity.
 
The purpose of this paper is to generalize the analysis of the determination of the P=parity of the $\Theta^+$-baryon, produced in the simplest reactions of the $NN$-interaction, $p+p\to \Theta^++\Sigma^+$ and $n+p\to \Theta^++\Lambda^0$, to the case of an arbitrary spin of the $\Theta^+$ hyperon. This analysis can be done in a model independent form, using the basic symmetry properties of the strong interaction, and does not need any specific dynamical assumptions for the above mentioned processes. This problem has been also analyzed in frame of a different formalism \cite{Uz04a}.
 
The central point of this analysis is based on the observation that there is a kinematical region near the reaction threshold, where the final baryons are produced in S-state. This region has an extension in the variable $Q$ ($Q=\sqrt{s}-M_1-M_2$, $s$ is the square of the total energy of the colliding nucleons, $M_1$ and $M_2$ are the masses of the produced hyperons), which is related to the finite radius of the strong  interaction. Note that, for the production of strange particles, in NN-collisions, such radius is expected to be of the order of $1/m_K$, $m_K$ is the kaon mass. Therefore, the corresponding threshold region, with S-wave production, is expected to quite wide.
 
This paper is organized as follows. In Section II we consider the production of 
$\Theta^+$-baryon with ${\cal J}^P=3/2^{\pm}$. We establish the spin structure of the threshold matrix elements for  $p+p\to \Theta^++\Sigma^+$ and $n+p\to \Theta^++\Lambda^0$, and analyze double spin polarization observables, sensitive to the parity of the $\Theta^+$-hyperon. The generalization on the case of any spin of the $\Theta^+$-hyperon is done in Section III.
\section{The reaction $N+N\to Y+\Theta^+$, $Y=\Lambda$ or $\Sigma$, 
${\cal J}^P(\Theta^+)=3/2^{\pm}$. }
The simplest reactions of $\Theta^+$ production in nucleon-nucleon collisions, 
$p+p\to \Sigma ^++\Theta^+$ and $n+p\to \Lambda ^0+\Theta^+$, with the lowest threshold energy  $E_L=$ 3.03 GeV (threshold momentum $p_L$=2.88 GeV) and 
2.82 GeV ($p_L$=2.66 GeV), respectively, and seem good candidates for the determination of the P-parity of the $\Theta^+$-hyperon, through the measurement of different polarization observables. The polarization transfer coefficient from the initial nucleon (beam or target) to the produced $Y$ hyperon, is relatively easy to measure, because the $\Lambda^0$ and the $\Sigma^+$-hyperons are self-analyzing particles.

Note, in this respect, that the DISTO collaboration  showed the feasibility of this method, by measuring the $D_{nn}$ coefficient at proton momentum of 3.67 GeV/c, which showed that {\it "$D_{nn}$ is large and negative ($\simeq -0.4$) over most of the kinematic region"} \cite{DISTO}. It was mentioned in \cite{Pa99} that a nonzero value of $D_{nn}$, in the threshold region, can be considered as the experimental confirmation of the pseudoscalar nature of the $K^+$ meson.

Polarization effects in $N+N\to Y+\Theta^+$ for ${\cal J}^P=1/2^{\pm}$, have been studied earlier \cite{Th03,Re04b,Uz04b,Ha03}, where it was shown that at least two observables, $A_{yy}$ and $D_{yy}$ are sensitive to the parity of the $\Theta^+$-hyperon, in different ways for $p+p\to \Sigma ^++\Theta^+$ and $n+p\to \Lambda ^0+\Theta^+$.

A similar derivation will be done for a more complicated case, ${\cal J}^P=3/2^{\pm}$, in the following sections.

\subsection{ \underline{The reaction $n+p\to \Lambda ^0+\Theta^+$, ${\cal J}^P(\Theta^+)=3/2^-$. }}
The selection rules with respect to the strong interaction allow a single threshold partial transition:
\begin{equation}
S_i=0,~\ell_i=1\to {\cal J}^P=1^-\to S_f=1,~\ell_f=0,
\label{eq:eq1}
\end{equation}
where $S_i$ and $\ell_i$ ($S_f$ and $\ell_f$) are the total spin and angular orbital momentum of the initial (final) baryons, ${\cal J}^P$ is the total angular momentum and P-parity of the colliding nucleons. We assume, all along this work, that the isotopic spin of the $\Theta^+$-hyperon is equal to zero.

Therefore the transition (\ref{eq:eq1}) results from the generalized Pauli principle. The corresponding matrix element can be written as:
\begin{equation}
{\cal M}^{(-)}_{\Lambda}=f^{(-)}_{\Lambda} (\tilde \chi_2 \sigma_y\chi_1)
( \chi_4^{\dagger} \sigma_y\tilde \chi_3^{\dagger}),~ \chi_4=\chi_a\hat k_a=\vec\chi \cdot\hat{\vec k},
\label{eq:mat}
\end{equation}
where $\hat{\vec k}$ is the unit vector along the three momentum of the colliding nucleons, in the reaction CM system, $\chi_1$ and $\chi_2$ are the two-component spinors of the initial nucleons, $\chi_3$ is the two-component spinor of the produced $\Lambda$ hyperon and $\chi_a$ is the two-component spinor with vector index $a$, describing the polarization properties of the $\Theta^+$ with spin 3/2, obeying to the following condition:
\begin{equation}
\vec\sigma\cdot\vec\chi =0,
\label{eq:mat}
\end{equation}
and, finally, $f^{(-)}_{\Lambda}$ is the partial amplitude for the singlet $np$ interaction.

Eq. (\ref{eq:mat}) allows to derive the dependence of the differential (and total) cross section on the polarizations $\vec P_1$ and $P_2$ of the colliding nucleons:
\begin{equation}
\displaystyle\frac{d\sigma}{d \Omega}^{(-)}(\vec P_1,\vec P_2)=\left (\displaystyle\frac{d\sigma}{d \Omega}\right )_0 (1-\vec P_1\cdot \vec P_2),
\label{eq:sig}
\end{equation}
independently on the amplitude $f^{(-)}_{\Lambda}$, i.e. independently on the concrete dynamics for the considered reaction.

It is straightforward to show that all polarization transfer coefficients, characterizing the dependence of the polarization of any final baryon on the polarization of any initial nucleon, vanish. The physical reason is that the singlet $np$-state, being 'closed' with respect to both nucleon spins, does not contain and does not transmit any information on the polarization to the final baryons.

Therefore the predictions:
$$A_{xx}=A_{yy}=A_{zz}=-1,$$
\begin{equation}
D_{xx}=D_{yy}=D_{zz}=0.
\label{eq:eqp}
\end{equation}
are typical properties for negative P-parity $\Theta^+$-production for $n+p\to \Lambda ^0+\Theta^+(3/2^-)$ in the threshold region.

\subsection{ \underline{The reaction $n+p\to \Lambda ^0+\Theta^+$, ${\cal J}^P(\Theta^+)=3/2^+$.}}

The symmetry selection rules (P-invariance, conservation of the total angular momentum and the validity of the generalized Pauli principle) result in the following three partial transitions:
\begin{eqnarray} 
S_i=1,~\ell_i=0&\to & {\cal J}^P=1^+\to S_f=1,~\ell_f=0,\nonumber\\
S_i=1,~\ell_i=2&\to & {\cal J}^P=1^+\to S_f=1,~\ell_f=0,\label{eq:eq5}\\
&&{\cal J}^P=2^+\to S_f=2,~\ell_f=0.\nonumber
\end{eqnarray}
The corresponding matrix element can be written as:
\begin{eqnarray} 
{\cal M}^{(+)}_{\Lambda}&=&f^{(+)}_{1\Lambda} (\tilde \chi_2 \sigma_y
\vec\sigma\cdot\hat{\vec k} \chi_1)( \vec\chi^{\dagger}\cdot\hat{\vec k}\sigma_y \tilde\chi_3^{\dagger})+\nonumber\\
&&f^{(+)}_{2\Lambda}\left [ \tilde \chi_2 \sigma_y(\sigma_m -\hat k_m
\vec\sigma\cdot \hat{\vec k})\chi_1 \right ] ( \chi_m^{\dagger} \sigma_y\tilde \chi_3^{\dagger})+,\label{eq:mpl}\\
&&if^{(+)}_{3\Lambda}\left [ \tilde \chi_2 
\sigma_y(\vec\sigma\times\hat{\vec k})_m\chi_1 \right ]
( \vec\chi^{\dagger}\cdot\hat{\vec k}\sigma_m\sigma_y \tilde\chi_3^{\dagger}),\nonumber
\end{eqnarray}
where $f^{(+)}_{i\Lambda}$, $i=1,2,3$ are the partial amplitudes for the case of positive parity of the $\Theta^+$ hyperon. Note that the amplitude 
$f^{(+)}_{1\Lambda}$ describes the triplet $np$-interaction in a state with zero value of total spin projection, but the amplitudes $f^{(+)}_{2,3\Lambda}$ describe the $np$ interaction with total spin projection equal to $\pm 1$.

Therefore, the cross section for polarized $\vec n+\vec p$-collisions can be written as:
\begin{equation}
\displaystyle\frac{d\sigma}{d \Omega}^{(+)}(\vec P_1\cdot \vec P_2)=\left (\displaystyle\frac{d\sigma}{d \Omega}\right )_0 
\left [1+{\cal A}_1^{(+)}\vec P_1\cdot \vec P_2 + {\cal A}_2^{(+)}\hat{\vec k}\cdot\vec P_1\hat{\vec k}\cdot\vec P_2\right ], 
\label{eq:sigpp}
\end{equation}
$$D^{(+)}_{\Lambda}{\cal A}_1^{(+)}=|f^{(+)}_{1\Lambda}|^2,~
D^{(+)}_{\Lambda}{\cal A}_2^{(+)}=2\left (-|f^{(+)}_{1\Lambda}|^2+|f^{(+)}_{2\Lambda}|^2
+|f^{(+)}_{3\Lambda}|^2\right ),
$$
where $D^{(+)}_{\Lambda}=|f^{(+)}_{1\Lambda}|^2+2\left (|f^{(+)}_{2\Lambda}|^2
+|f^{(+)}_{3\Lambda}|^2\right )$.

The relation $3{\cal A}_1^{(+)}+{\cal A}_2^{(+)}=1$, which is correct for any values of partial amplitudes $f^{(+)}_{i\Lambda}$, results from the absence of the singlet $np$-interaction - for the considered case.

One can find from (\ref{eq:sigpp}):
\begin{equation}
A_{xx}^{(+)}=A_{yy}^{(+)}=\displaystyle\frac{1}{2}\left (1-A_{zz}^{(+)}\right )=
\displaystyle\frac{|f^{(+)}_{1\Lambda}|^2}{|f^{(+)}_{1\Lambda}|^2+
2\left (|f^{(+)}_{2\Lambda}|^2
+|f^{(+)}_{3\Lambda}|^2\right )}\ge 0,
\label{eq:sigaa}
\end{equation}
which is different from the case of negative P-parity. 

The same conclusion holds for the case of the coefficients of polarization transfer. The dependence of the $\Lambda$-polarization $\vec P_{\Lambda}$ on the initial beam polarization, $\vec P$ (described by the two-component spinor $\chi_2$) can be written in the following form, which holds for S-wave production:

\begin{equation} 
\vec P_{\Lambda}=p_1^{(+)}\vec P + p_2^{(+)}\hat{\vec k}(\hat{\vec k}\cdot\vec P),
\label{eq:eq10}
\end{equation}
where $ p_{1,2}$ are real coefficients, which can be expressed as a function of the amplitudes as: 
$$D^{(+)}_{\Lambda}p_1^{(+)}=2|f^{(+)}_{2\Lambda}|^2-Re f^{(+)}_{1\Lambda} (f^{(+)}_{2\Lambda}+2 f^{(+)}_{3\Lambda})^*, 
$$
\begin{equation} 
D^{(+)}_{\Lambda}p_2^{(+)}=-|f^{(+)}_{2\Lambda}|^2+2|f^{(+)}_{3\Lambda}|^2+Re
\left [ f^{(+)}_{1\Lambda}f^{(+)*}_{2\Lambda}+2(f^{(+)}_{1\Lambda}+f^{(+)}_{2\Lambda})f^{(+)*}_{3\Lambda}
\right ].
\label{eq:eq11}
\end{equation}
Therefore, generally, the polarization transfer coefficients:
$$
{\cal D}^{(+)}_{xx}={\cal D}^{(+)}_{yy}=p_1^{(+)},~{\cal D}^{(+)}_{zz}=p_1^{(+)}+p_2^{(+)} $$
do not vanish, in this case.

\begin{figure}
\mbox{\epsfxsize=15.cm\leavevmode \epsffile{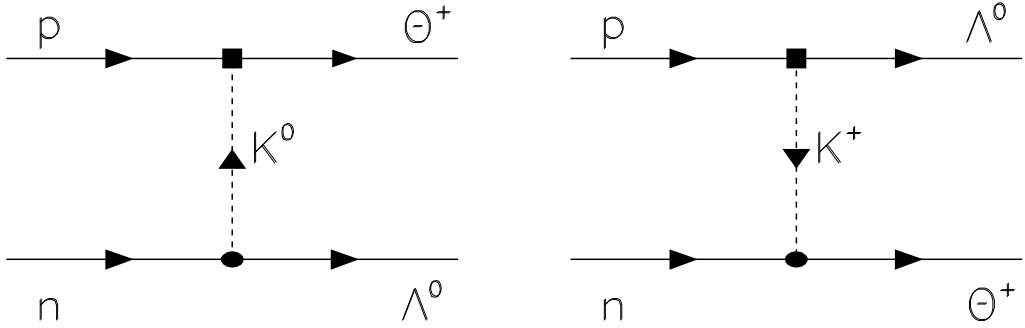}}
\caption{ $K$-exchange for the reaction $ n+p \to \Lambda^0 +\Theta^+$.
}
\label{fig:fig1}
\end{figure}

To do a numerical estimation of these polarization observables, let us consider a simple model for 
$n+p\to \Lambda+\Theta^+$, based on $t$-channel $K$-exchange (Fig. \ref{fig:fig1}).
Note that only the coherent sum of the two diagrams, with vertices which satisfy the isotopic invariance, results a correct spin structure of the threshold matrix element, see Eq. (\ref{eq:mpl}). Each diagram generate a spin structure which includes a singlet amplitude, which cancels in the sum. The final result for the considered mechanism can be written as:
\begin{equation}
f^{(+)}_{1\Lambda}=-f^{(+)}_{3\Lambda},~f^{(+)}_{2\Lambda}=0.
\label{eq:eq12}
\end{equation}
Let us stress that these predictions do not depend on the values of the coupling constants $g_{N\Lambda K}$ and $g_{N\Theta K}$, and on the phenomenological form factors which are important ingredients of this model. These quantities enter in the calculation of the absolute values of the differential (and total) cross section, but not in the predictions for the polarization observables.

Substituting  Eq. (\ref{eq:eq12}) in Eqs. (\ref{eq:sigpp}) and (\ref{eq:eq11}), one can find:
\begin{equation}
A_{xx}^{(+)}=A_{yy}^{(+)}=A_{zz}^{(+)}=\displaystyle\frac{1}{3},~
D_{xx}^{(+)}=D_{yy}^{(+)}=-\displaystyle\frac{2}{3},~D_{zz}^{(+)}=\displaystyle\frac{2}{3}.
\label{eq:eq13}
\end{equation}
Note that the predicted result $D_{xx}^{(+)}=-2/3$ coincides with $D_{nn}^{(+)}=-2/3$, which can be found for $\vec p+p\to \vec\Lambda +K^++p$ - in framework of a similar model, based on $K$-exchange, in agreement with the existing DISTO results. 

Another confirmation of the validity of this simple model, which takes into account only $K$-exchange for threshold strange particle production in $pp$-collisions is the large value of the ratio 
$\sigma(pp\to \Lambda K^+ p)/\sigma(pp\to \Sigma K^+ p)$, measured in COSY \cite{Mo02}.

Considering the similarity of final states in $N+N\to \Lambda +K^+ +N$ and in the reaction of interest here, $n+p\to  \Lambda +\Theta^+\to\Lambda +K^+ +n$, one can assume that the $K$-exchange model can be applied here also. Of course, this is not a proof, but a qualitative argument to justify the model taken here for a quick estimation of polarization effects in $\Lambda +\Theta^+ (3/2^+)$, with a complicated spin structure.

The result found here, $D_{yy}^{(+)}=-2/3$, (which is model dependent in case of ${\cal J}^P(\Theta^+)=3/2^+$ ) is very far from the value $D_{yy}^{(-)}=0$, which has been found in model independent way for the opposite parity of  $\Theta^+$, and shows the level of accuracy which will be necessary, in order to discriminate the different parities of the $\Theta^+$-baryon.
\subsection{ \underline{The reaction $p+p\to \Sigma ^++\Theta^+,~{\cal J}^P(\Theta^+)=3/2^+$.}}
For  a positive parity, only one spin transition is allowed at threshold:
\begin{equation}
S_i=0,~\ell_i=2\to {\cal J}^P=2^+\to S_f=2,~\ell_f=0,
\label{eq:eqp}
\end{equation}
to which corresponds the following matrix element:
\begin{equation}
{\cal M}^{(+)}_{\Sigma }=f^{(+)}_{\Sigma } (\tilde \chi_2 \sigma_y\chi_1)
( \vec\chi^{\dagger} \cdot\hat{\vec k} \vec\sigma\cdot\hat{\vec k}\sigma_y\tilde \chi_3^{\dagger}).
\label{eq:mat}
\end{equation}
The dependence of the differential cross section on the polarizations of the initial nucleons is:
$$
\displaystyle\frac{d\sigma}{d \Omega}(\vec p\vec p\to \Sigma ^+\Theta^+)=
\left (\displaystyle\frac{d\sigma}{d \Omega}\right )_0 (1-\vec P_1\cdot \vec P_2),
$$
and the polarization transfer coefficients:
\begin{equation}
D_{xx}^{(+)}(\Sigma )=D_{yy}^{(+)}(\Sigma )=D_{zz}^{(+)}(\Sigma )=0.
\label{eq:eq16p}
\end{equation}
i.e. the result is similar to the reaction $n+p\to \Lambda^0+\Theta^+(3/2^-)$ but with opposite parity!

\subsection{ \underline{The reaction $p+p\to \Sigma ^++\Theta^+$, ${\cal J}^P(\Theta^+)=3/2^-$.}}

The selection rules related to the strong interaction allow the following partial transitions:
\begin{eqnarray} 
S_i=1,~\ell_i=1&\to & {\cal J}^P=1^-\to S_f=1,~\ell_f=0,\nonumber\\
&\to & {\cal J}^P=2^-\to S_f=2,~\ell_f=0,\label{eq:eq17}\\
S_i=1,~\ell_i=3&\to &{\cal J}^P=2^-\to S_f=2,~\ell_f=0.\nonumber
\end{eqnarray}
with matrix element:
\begin{eqnarray} 
{\cal M}^{(-)}_{\Sigma}&=&f^{(-)}_{1\Sigma} (\tilde \chi_2 \sigma_y
\vec\sigma\cdot\hat{\vec k} \chi_1)( \vec\chi^{\dagger}\cdot\hat{\vec k}
\vec\sigma\cdot\hat{\vec k}\sigma_y \tilde\chi_3^{\dagger})+\nonumber\\
&&f^{(-)}_{2\Sigma}\left [ \tilde \chi_2 \sigma_y(\sigma_m -\hat k_m
\vec\sigma\cdot \hat{\vec k})\chi_1 \right ] (\vec \chi^{\dagger}\cdot\hat{\vec k}\sigma_m \sigma_y\tilde \chi_3^{\dagger})+,\label{eq:mpl1}\\
&&if^{(+)}_{3\Sigma}\left [ \tilde \chi_2 
\sigma_y(\vec\sigma\times\hat{\vec k})_m\chi_1 \right ]
( \chi_{m}^{\dagger}\sigma_y \tilde\chi_3^{\dagger}),\nonumber
\end{eqnarray}
The dependence of the differential cross section on the polarizations of the colliding nucleons can be written in a standard form, which holds for S-wave production:
\begin{equation}
\displaystyle\frac{d\sigma}{d \Omega}^{(-)}(\vec p\vec p\to \Sigma ^+\Theta^+)=
\left (\displaystyle\frac{d\sigma}{d \Omega}\right )_0 
\left [1+{\cal A}_{1\Sigma}^{(-)}\vec P_1\cdot \vec P_2 + {\cal A}_{2\Sigma}^{(-)}
\hat{\vec k}\cdot\vec P_1\hat{\vec k}\cdot\vec P_2.\right ]
\label{eq:eq19}
\end{equation}
with the following formulas for ${\cal A}_{1,2\Sigma}$:
\begin{equation}
D_{\Sigma}^{(-)}{\cal A}_{1\Sigma}^{(-)}=|f^{(+)}_{1\Sigma}|^2,
~ D_{\Sigma}^{(-)}{\cal A}_{2\Sigma}^{(-)}=2\left [ -|f^{(+)}_{1\Sigma}|^2+|f^{(+)}_{2\Sigma}|^2+|f^{(+)}_{3\Sigma}|^2\right ], 
\label{eq:eq20}
\end{equation}
with 
$$ 
D_{\Sigma}^{(-)}=|f^{(-)}_{1\Sigma}|^2+2\left (|f^{(-)}_{2\Sigma}|^2+|f^{(-)}_{3\Sigma}|^2\right ),
$$
i.e.
\begin{equation}
A_{xx}^{(-)}(\Sigma)=A_{yy}^{(-)}(\Sigma)={\cal A}_{1\Sigma}^{(-)}\ge 0.
\label{eq:eq21}
\end{equation}
The polarization transfer coefficients can be expressed as a function of the amplitudes as:
\begin{equation} 
\vec P_{\Sigma}^{(-)}=p_{1\Sigma}^{(-)}\vec P + p_{2\Sigma}^{(-)}\hat{\vec k}(\hat{\vec k}\cdot\vec P),
\label{eq:eq22}
\end{equation}
\begin{equation}
D_{\Sigma}^{(-)}p_{1\Sigma}^{(-)}=Ref^{(-)}_{1\Sigma}(2f^{(-)}_{2\Sigma}+ f^{(-)}_{3\Sigma})^*,
\label{eq:eq23}
\end{equation}
\begin{equation}
D_{\Sigma}^{(-)}p_{2\Sigma}^{(-)}=2|f^{(-)}_{2\Sigma}|^2- |f^{(-)}_{3\Sigma}|^2 - Re~f^{(-)}_{1\Sigma}f^{(-)*}_{3\Sigma} -2Re~ (2f^{(-)}_{1\Sigma}- f^{(-)}_{3\Sigma})f^{(-)*}_{2\Sigma}.
\label{eq:eq24}
\end{equation}
Let us evaluate these observables again in framework of the above mentioned $K$-exchange model for $p+p\to\Sigma ^++\Theta^+ $. Two Feynman diagrams, in analogy with those of Fig. \ref{fig:fig1} generate the following relation between the partial amplitudes $f^{(-)}_{i\Sigma}$:
\begin{equation}
f^{(-)}_{2\Sigma}=0,~f^{(-)}_{1\Sigma}=-f^{(-)}_{3\Sigma},
\label{eq:eq25}
\end{equation}
again independent on coupling constants and phenomenological form factors. This allows to find:
\begin{equation}
A_{xx}^{(-)}(\Sigma)=A_{yy}^{(-)}(\Sigma)=A_{zz}^{(-)}(\Sigma)=\displaystyle\frac{1}{3},~
D_{xx}^{(-)}(\Sigma)=D_{yy}^{(-)}(\Sigma)=D_{zz}(\Sigma)^{(-)}=-\displaystyle\frac{2}{3}.
\label{eq:eq26}
\end{equation}

\section{\underline{The reaction $ N+N\to Y+\Theta^+(j^P) , ~j^P$ is any}}

In this section, we consider the case of any $j^P$ for the $\Theta^+$ hyperon, produced in the reactions $N+N\to Y+\Theta^+$, $Y=\Lambda$ or $\Sigma^+$-hyperon.

At the reaction threshold, the polarization properties of the $\Theta^+(j^P)$ can be described by the non-relativistic limit of the corresponding Rarita-Schwinger spinor, more exactly by the following two component spinor $\chi_{a_1,...a_n},$ $n=j-1/2,j\ge 3/2$, with definite number of vector indices. Such spinor has to satisfy the conditions:
\begin{equation}
\sigma_a\chi_{aa_2,...a_n}=0, ~ \chi_{aaa_3...a_n}=0, 
\label{eq:eq27}
\end{equation}
$$\chi_{a_1a_2...a_n}=\chi_{a_2a_1...a_n},$$
this last property guarantees the symmetry with respect to the interchange of any pair of vector indices $a_i$, $i=1,..,n$.

Evidently such construction has $2j+1$ independent components, as it must be the case for a particle with spin $j$.

We will apply this general formalism to the process $ p+p \to \Sigma^+ +\Theta^+$ in threshold conditions. The process $ n+p \to \Lambda^0 +\Theta^+$ can be considered in a similar way.
\subsection{\underline{The reaction $p+p\to \Sigma ^+ + \Theta ^+ (j^+)$}}
In this section we consider the case when the P-parity of $\Theta^+$ is positive.  the selection rules allow only one partial transition:
\begin{equation}
S_i=0,~\ell_i=\mbox{even~}\to {\cal J}=\ell\to S_f=\ell_i,~\ell_f=0,
\label{eq:eq28}
\end{equation}
with the following relation between $\ell_i$ and $j$:
\begin{eqnarray} 
\ell_i&=&j-\displaystyle\frac{1}{2},~\mbox{if~} P(\Theta)=(-1)^{j-1/2},~\mbox{i.e.~}~j^P=\displaystyle\frac{1}{2}^+,~
\displaystyle\frac{5}{2}^+,~\displaystyle\frac{9}{2}^+...,\nonumber\\
\ell_i&=&j+\displaystyle\frac{1}{2},~\mbox{if~} P(\Theta)=(-1)^{j+1/2},~\mbox{i.e.~}~j^P=\displaystyle\frac{3}{2}^+,~
\displaystyle\frac{7}{2}^+,~\displaystyle\frac{11}{2}^+...,\label{eq:eq29}.
\end{eqnarray}
The first case corresponds to natural $(n)$ parity and the second to unnatural $(u)$ parity.

The corresponding matrix element can be written as follows:
\begin{equation}
{\cal M}^{(+)}=f^{(+)}(j) (\tilde \chi_2 \sigma_y\chi_1)(\chi^{\dagger}_4 \sigma_y\tilde \chi_3^{\dagger}).
\label{eq:eq30}
\end{equation}
with 
\begin{eqnarray*} 
\chi_4&=&\chi_{a_1a_2,...a_n}\hat k_{a_1}\hat k_{a_2}...\hat k_{a_n}\mbox{for~natural P-parity},\\
\chi_4&=&\vec\sigma\cdot\hat{\vec k} \chi_{a_1a_2,...a_n}\hat k_{a_1}\hat k_{a_2}...\hat k_{a_n}\mbox{for~unnatural P-parity},
\end{eqnarray*}
where $f^{(+)}(j)$ is the partial amplitude, depending on the $\Theta$-spin $j$.

Independently on the value of the $\Theta^+$ spin $j$, from the matrix element (\ref{eq:eq30})  one derives the general form for all double-spin polarization observables:
$$
\displaystyle\frac{d\sigma}{d \Omega}^{(+)}(j)=\left (\displaystyle\frac{d\sigma}{d \Omega}\right )_0 (1-\vec P_1\cdot \vec P_2),$$
\begin{equation}
D_{xx}^{(+)}(j)=D_{yy}^{(+)}(j)=D_{zz}^{(+)}(j)=0,
\label{eq:eqp}
\end{equation}
due to the 'closed' initial singlet $pp$-state.
\subsection{ \underline{The reaction $p+p\to \Sigma ^++\Theta^+(j^-)$}}

The case of negative P-parity must be treated separately, for natural and unnatural parity states.

\subsubsection{ Natural P-parity}

For $j^P=\displaystyle\frac{3}{2}^-,~\displaystyle\frac{7}{2}^-,...$ the following partial transitions are allowed:
\begin{eqnarray} 
S_i=1,~\ell_i=j-\displaystyle\frac{1}{2}&\to & {\cal J}=j-\displaystyle\frac{1}{2}\to S_f=j-\displaystyle\frac{1}{2},\nonumber\\
&\to & {\cal J}=j+\displaystyle\frac{1}{2}\to S_f=j+\displaystyle\frac{1}{2},\label{eq:eq32}\\
S_i=1,~\ell_i=j+\displaystyle\frac{3}{2}&\to &{\cal J}=j+\displaystyle\frac{1}{2}\to S_f=j+\displaystyle\frac{1}{2},\nonumber
\end{eqnarray}
with the corresponding spin structure of the matrix element:
\begin{eqnarray} 
{\cal M}^{(-)}_n(j)&=&f^{(-)}_{1n}(j)(\tilde \chi_2 \sigma_y
\vec\sigma\cdot\hat{\vec k} \chi_1)( \chi^{\dagger}_4\vec\sigma\cdot\hat{\vec k}
\sigma_y \tilde\chi_3^{\dagger})+\nonumber\\
&&if^{(-)}_{2n}(j)\left [ \tilde \chi_2 
\sigma_y(\vec\sigma\times\hat{\vec k})_m\chi_1 \right ]
( \chi_{4m}^{\dagger}\sigma_y \tilde\chi_3^{\dagger})+,\label{eq:eq33}\\
&&f^{(-)}_{3n}(j)( \tilde \chi_2 
\sigma_y\sigma_m\chi_1)\left [ \chi^{\dagger}_4(\sigma_m -\hat k_m
\vec\sigma\cdot \hat{\vec k})\sigma_y\tilde\chi_3^{\dagger}\right ],\nonumber
\end{eqnarray}
where $\chi_4\equiv \chi_{a_1a_2,...a_n}\hat k_{a_1}\hat k_{a_2}...\hat k_{a_n}$, 
$\chi_{4m}\equiv \chi_{ma_2,...a_n}\hat k_{a_2}...\hat k_{a_n}$, and $f^{(-)}_{in}$, $i=1-3$ are the independent partial amplitudes.

\subsubsection{ Unnatural P-parity}

The case  $j^P=\displaystyle\frac{1}{2}^-,~\displaystyle\frac{5}{2}^-,...$ generates the following set of triplet transitions:
\begin{eqnarray} 
S_i=1,~\ell_i=j-\displaystyle\frac{3}{2}&\to & {\cal J}=j-\displaystyle\frac{1}{2}\to S_f=j-\displaystyle\frac{1}{2},\nonumber\\
S_i=1,~\ell_i=j+\displaystyle\frac{1}{2}&\to & {\cal J}=j-\displaystyle\frac{1}{2}\to S_f=j-\displaystyle\frac{1}{2},\label{eq:eq34}\\
&\to &{\cal J}=j+\displaystyle\frac{1}{2}\to S_f=j+\displaystyle\frac{1}{2},\nonumber
\end{eqnarray}
with corresponding matrix element:
\begin{eqnarray} 
{\cal M}^{(-)}_u(j)&=&f^{(-)}_{1u}(j)
(\tilde \chi_2 \sigma_y\vec\sigma\cdot\hat{\vec k} \chi_1)
(\chi^{\dagger}_4\sigma_y \tilde\chi_3^{\dagger})+\nonumber\\
&&if^{(-)}_{2u}(j)\left [ \tilde \chi_2 
(\vec\sigma\times\hat{\vec k})_m\chi_1 \right ]
( \chi_4^{\dagger}\sigma_m\sigma_y \tilde\chi_3^{\dagger})+,\label{eq:eq35}\\
&&f^{(-)}_{3n}(j)\left [ \tilde \chi_2 (\sigma_m -\hat k_m
\vec\sigma\cdot \hat{\vec k}) \chi_1\right ]
(\chi^{\dagger}_{4m}\sigma_y\tilde\chi_3^{\dagger}).\nonumber
\end{eqnarray}
Eqs. (\ref{eq:eq33}) and (\ref{eq:eq35})  allow to express the coefficients ${\cal A}^{(-)}(j)$ in terms of the partial amplitudes $f^{(-)}_{in}(j)$, or ($f^{(-)}_{iu}(j)$):
$$
{\cal A}_1^{(-)}(j)= \displaystyle\frac{|f^{(-)}_1(j)|^2}{|f^{(-)}_1(j)|^2
+2(|f^{(-)}_2(j)|^2+|f^{(-)}_3(j)|^2},$$
\begin{equation}
3{\cal A}_1^{(-)}(j)-{\cal A}_2^{(-)}(j)=1,
\label{eq:eq36}
\end{equation}
(the indices $u$ and $n$ are not indicated in these formulas).

One can see that for any $j$, the negative parity of the $\Theta$-hyperon results in:
\begin{equation}
A_{xx}^{(-)}(j)=A_{yy}^{(-)}(j)\ge 0.
\label{eq:eq37}
\end{equation}
In the general case, the numerical value of the asymmetries depend on $j$, being, however, positive. This holds for the process $ p+p\to \Sigma ^++\Theta^+$.

Note, finally, that in the $K$-meson exchange model (which is taken here for illustrative purposes), one can find:
\begin{equation}
f^{(-)}_2(j)=0,~f^{(-)}_1(j)=\pm f^{(-)}_3(j)\ne 0,
\label{eq:eq38}
\end{equation}
where the sign $\pm$ corresponds to natural or unnatural P-parity, with the following universal result for any $j$:
\begin{equation}
A_{xx}^{(-)}(j)=A_{yy}^{(-)}(j)=A_{zz}^{(-)}(j)=\displaystyle\frac{1}{3},
\label{eq:eq37}
\end{equation}
i.e. in this model the asymmetries do not depend on $j$.
\section{Conclusions}
We calculated double spin polarization observables in  simplest processes of $\Theta^+$ production in $NN$ collisions, $n+p\to \Lambda ^0+\Theta^+$ and $p+p\to \Sigma ^++\Theta^+$.
We proved that the spin correlation coefficients $A_{xx}$ and  $A_{yy}$ (in the collisions of transversally polarized nucleons) and the polarization transfer coefficient $D_{yy}$ (characterizing the transversal polarization of the final hyperon, $\Sigma ^+$ or $\Lambda ^+$ -self analyzing particles, emitted in the collision of a polarized (unpolarized) nucleon beam with unpolarized (polarized) nucleon target) can be considered as model independent filters for the determination of the P-parity of the $\Theta^+$ hyperon, whatever value takes its spin.

We found that the spin structure of the matrix element is essentially different for different parities. Whereas only the singlet amplitude is present in $p+p\to \Sigma ^++\Theta^+$, for positive parity and any spin, for the process $n+p\to \Lambda^0+\Theta^+$ the singlet amplitude is associated to a negative parity ( for any spin).

In the same formalism, we established that the process $p+p\to \Sigma ^++\Theta^+(j^-)$ is characterized by three triplet amplitudes, for any $j$, $j\ge 3/2$. Similar triplet amplitudes define the spin structure of the threshold matrix element for $n+p\to \Lambda^0+\Theta^+$, in case of positive $\Theta^+$ parity.

We stress once more that the functional forms of the spin structure and  polarization phenomena in both reactions: $n+p\to \Lambda^0+\Theta^+$ and $p+p\to \Sigma ^++\Theta^+$, has been done in model independent way, using only the general selection rules which hold for strong interaction, as conservation of total angular momentum, isotopic invariance, Pauli principle.
 
This allowed us to generalize the previously proposed methods for the $\Theta^+$ parity determination in the case of $j=1/2$ to any value of the spin. The main result of this work can be formulated as follows: the sign of the asymmetry $A_{yy}$ (which is different for $p+p\to \Sigma ^++\Theta^+$ and $n+p\to \Lambda^0+\Theta^+$) is uniquely related to the $\Theta^+$ parity, independently on its spin $j$. This is also correct for the polarization transfer coefficient $D_{yy}$: a value equal or different from zero of this coefficient is an unambiguous signature of the discussed P-parity.

Numerical estimations of these polarization observables were done using a simple but realistic dynamical model, for the considered reactions, based on $K$-meson exchange. In framework of this model, the relations among the threshold partial amplitudes are independent on the coupling constants for the vertices of the considered Feynman diagrams and on the parametrization of the phenomenological form factors, quantities which enter in the calculation of the differential cross section. Polarization phenomena do not depend on these ingredients of the model, but are more sensitive to more general properties of the reaction mechanism, such as the quantum numbers of the exchanged particles.

The polarization phenomena discussed here are T-even, so they do not vanish even in the framework of a simple model, with real amplitudes. Moreover they are not very sensitive to the effects of initial and final state interaction.

{}

\end{document}